\documentclass[3p,times]{elsarticle}

\usepackage{ecrc}

\usepackage{epstopdf}

\volume{00}

\firstpage{1}

\journalname{Nuclear Physics A}

\runauth{J.F. Grosse-Oetringhaus for the ALICE collaboration}

\jid{nupha}

\jnltitlelogo{Nuclear Physics A}

\usepackage{graphicx}
\usepackage{amsmath,amssymb}

\usepackage{lineno}

\usepackage{units}

\biboptions{sort&compress}

\newcommand{\gevc}{GeV/\emph{c}}
\newcommand{\pt}{p_{\rm T}}
\newcommand{\expval}[1]{\langle #1 \rangle}

\begin{document}

\begin{frontmatter}



\title{Overview of ALICE Results at Quark Matter 2014}

\author{Jan Fiete Grosse-Oetringhaus (for the ALICE\fnref{col1} Collaboration)}
\fntext[col1] {A list of members of the ALICE Collaboration and acknowledgements can be found at the end of this issue.}
\address{CERN, 1211 Geneva 23, Switzerland}

\begin{abstract}
The results released by the ALICE collaboration at Quark Matter 2014 address topics from identified-particle jet fragmentation functions in pp collisions, to the search for collective signatures in p--Pb collisions to precision measurements of jet quenching with D mesons in Pb--Pb collisions.
This paper gives an overview of the contributions (31 parallel talks, 2 flash talks and 80 posters) by the ALICE collaboration at Quark Matter 2014.
\end{abstract}


\end{frontmatter}


\section{Introduction}

The ALICE collaboration has presented numerous new results at Quark Matter 2014 which advance our knowledge about the dynamics of ultrarelativistic proton and heavy-ion collisions in various directions. Among the main topics are precision measurements of jet quenching in heavy-ion collisions as well as further understanding of
possible collectivity in proton--lead collisions. These proceedings can only present a subset of the results, for more information the reader is referred to \cite{aiola, alexandre, andrei, arnaldi, bailhache, bellini, belmont, bjelogrlic, blanco, bockf, bookj, castillo, connors, dobrin, festanti, gangadharan, gunji, knichel, kohler, lakomov, li, loggins, lu, marin, martin, milano, nystrand, siddhanta, russo, timmins, toia, zhang, zhou}.
These discuss light flavour \cite{alexandre, andrei, bellini, knichel, martin}, heavy flavour \cite{bailhache, festanti, li, russo, bjelogrlic}, jet \cite{aiola, connors, lu, zhang}, photon \cite{bockf, marin}, quarkonia \cite{arnaldi, blanco, castillo, bookj, lakomov}, dielectron \cite{kohler} and correlation \cite{belmont, gangadharan, dobrin, loggins, milano, timmins, zhou} measurements as well as ultra-peripheral collisions \cite{nystrand}, event characterization \cite{toia} and the ALICE upgrade \cite{siddhanta, gunji}.

\section{Results from Proton--Lead Collisions}

Proton--lead collisions at the LHC, initially thought of just as a control experiment for Pb--Pb collisions, have triggered significant interest when it became clear that non-trivial (potentially final-state) effects were present as compared to an incoherent superposition of nucleon--nucleon collisions. In particular, the behavior of the $\langle \pt \rangle$ as a function of multiplicity~\cite{meanpt}, the multiplicity dependence of identified-particle spectra~\cite{spectra} and the discovery of long-range correlation structures~\cite{doubleridge} as well as their dependence on particle type~\cite{doubleridgepid} should be noted. These structures are typically found in heavy-ion collisions where they are interpreted as a signature of collectivity.
A new measurement using four-particle cumulants has established that higher-order correlations contribute to these long-range correlations~\cite{timmins,1406.2474}.

\subsection{Nuclear Modification Factor}

\begin{figure}[t!]
    \begin{center}
        \includegraphics*[width=0.54\textwidth]{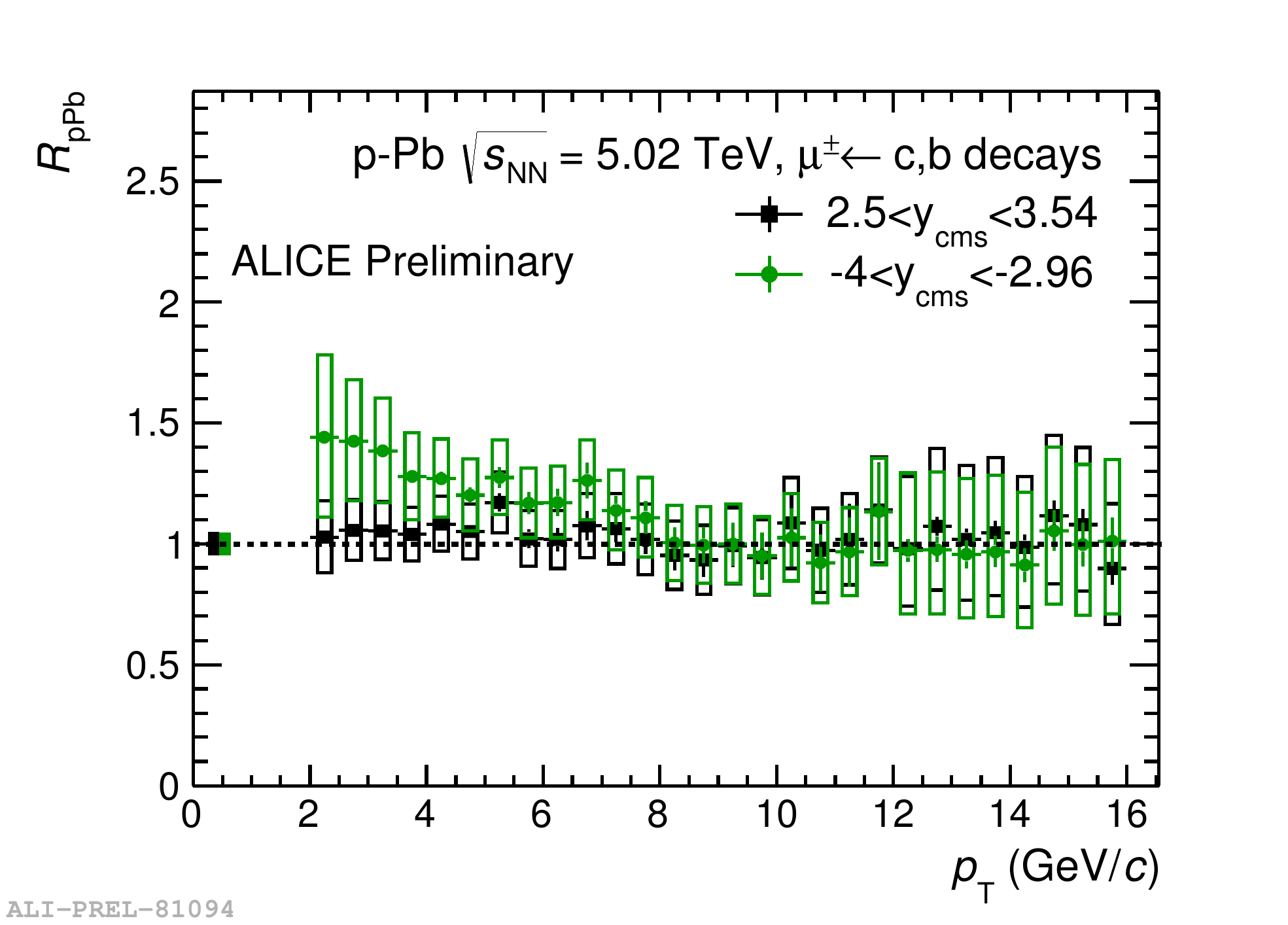}
        \hfill
        \includegraphics*[width=0.45\textwidth]{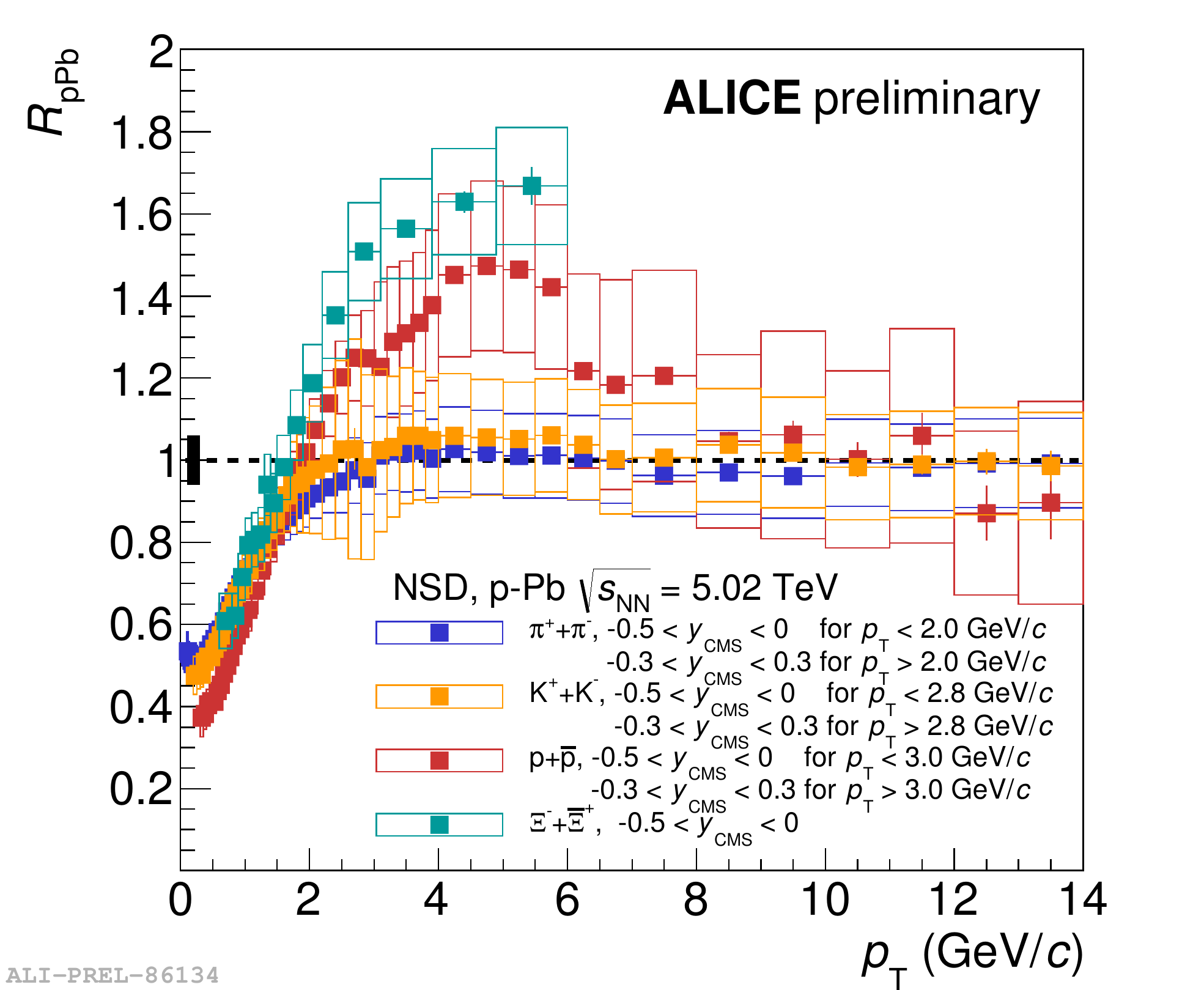}
        \caption{\label{rpa} Nuclear modification factor $R_{\rm pPb}$ of forward muons from heavy-flavour decays measured in p-going (black) and Pb-going (green) directions (left panel) and of $\pi$ (blue), K (orange), p (red) and $\Xi$ (green) measured at mid-rapidity (right panel).}
    \end{center}
\end{figure}

While the observations discussed in the previous paragraph hint at novel effects in p--Pb collisions, observables sensitive to energy loss in a hot and dense medium show no significant deviation from an incoherent superposition of nucleon--nucleon collisions. This can be seen for example in measurements of the nuclear-modification factor $R_{\rm pPb}$ defined as the ratio of $\pt$ spectra in p--Pb collisions and in pp collisions scaled by the number of nucleon--nucleon collisions. Results from minimum-bias collisions for charged particles up to \unit[50]{\gevc}~\cite{knichel,1405.2737}, for jets up to \unit[90]{\gevc}~\cite{aiola, connors}, for D mesons up to about \unit[20]{\gevc}~\cite{russo,1405.3452}, for muons from heavy-flavour decays measured at forward rapidities up to \unit[16]{\gevc}~\cite{li} as well as for electrons from beauty decays up to about \unit[6]{\gevc} show no deviation from unity at high $\pt$. As an example the nuclear modification factor of muons measured at forward rapidities is shown in the left panel of Fig.~\ref{rpa} for p-going and Pb-going directions. No significant energy loss seems to be at play in minimum-bias p--Pb collisions. The centrality dependence of these effects and the related difficulties encountered
are discussed further below (Section~\ref{sec_centrality}).

The minimum-bias nuclear-modification factor for charged particles shows an increase up to about 1.1 at \unit[3--4]{\gevc}. It is interesting to note that such an enhancement has already been observed at lower energies and is referred to as Cronin effect~\cite{cronin}. This peak shows a strong particle-type dependence~\cite{knichel}, presented in the right panel of Fig.~\ref{rpa}: it is not observed for $\pi$ and K, while it is rather strong for p and $\Xi$ where $R_{\rm pPb}$ reaches values of 1.5--1.7. Such a behavior has been interpreted as a consequence of radial flow which alters the $\pt$ spectra. This can be further understood by analyzing the particle spectra within blast-wave models which are a proxy for hydrodynamic modeling. Within these models one finds that the extracted average radial-flow expansion velocity $\expval{\beta_{\rm T}}$ is larger in p--Pb collisions than in Pb--Pb collision at the same multiplicity~\cite{andrei}. This can be interpreted as indication that a larger density gradient is prevailing in p--Pb collisions. However, applying blast-wave fits to pp collisions, large values of $\expval{\beta_{\rm T}}$ are also found in measured pp events and PYTHIA~\cite{pythia} generated pp events underlining that there may be other mechanisms than the expansion of a hot and dense fireball producing signatures compatible with radial expansion.

\subsection{Jet Modification at low $\pt$}

\begin{figure}[t!]
    \begin{center}
        \includegraphics*[width=0.43\textwidth]{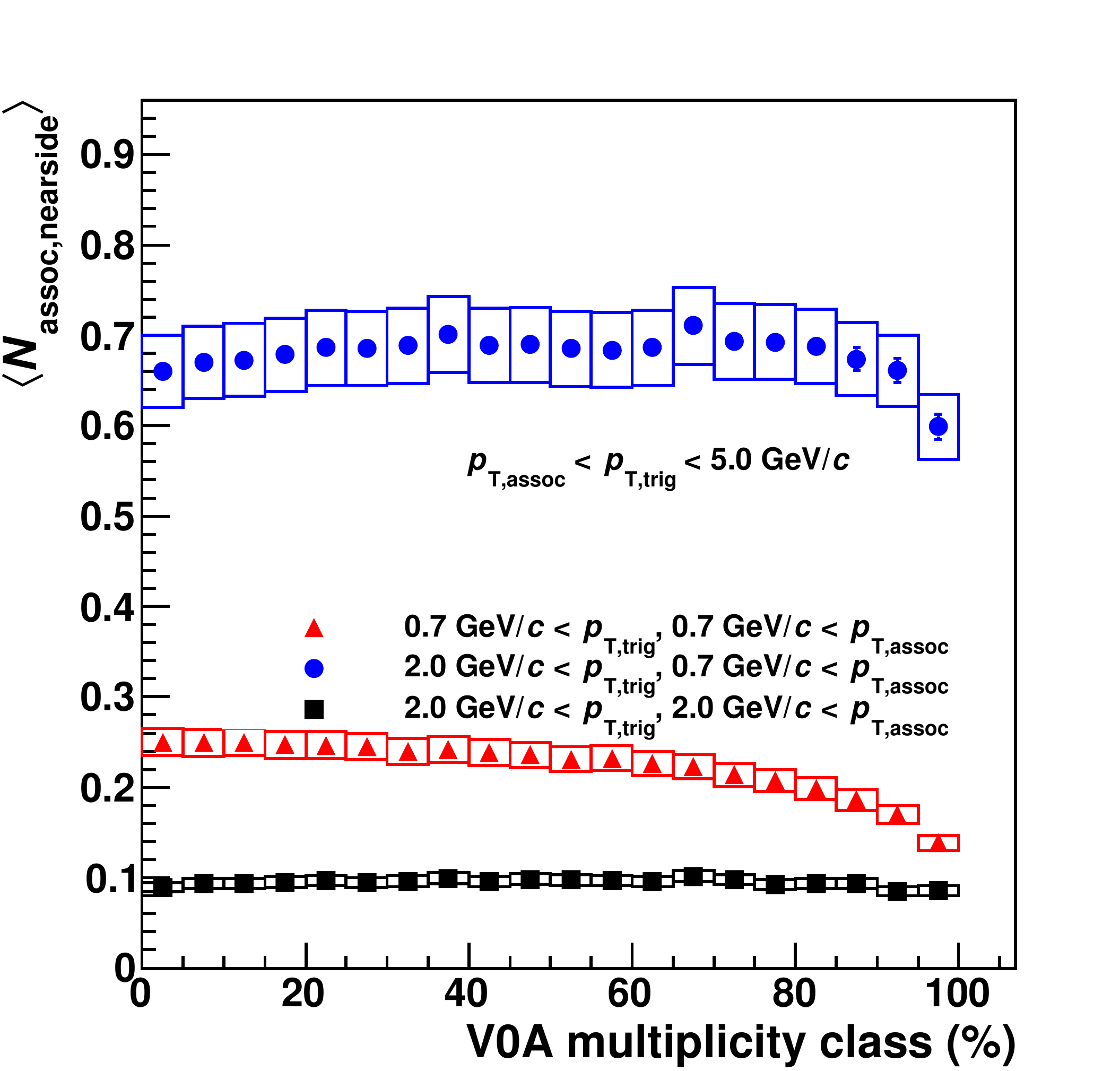}
        \hfill
        \includegraphics*[width=0.56\textwidth]{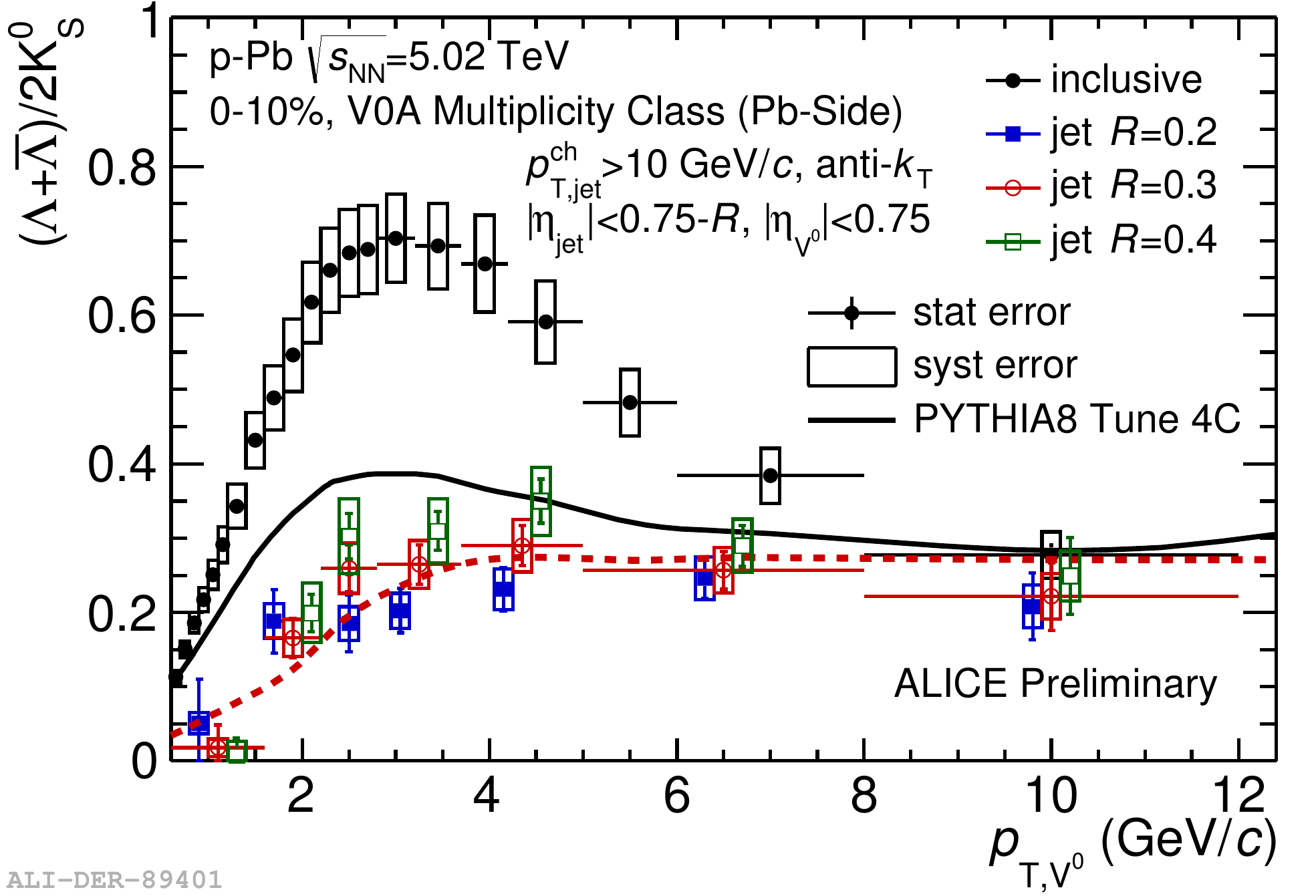}
        \caption{\label{minijets_lk0ratio} Left panel: Near-side jet-like associated yield as a function of multiplicity class (measured with the forward V0A detector, $2.8 < \eta_{\rm lab} < 5.1$) after subtraction of the long-range contribution for several $\pt$ ranges~\cite{1406.5463}. \newline Right panel: $\Lambda$ to K$^{\rm 0}_{\rm S}$ ratio as a function of their $\pt$ (denoted with $p_{\rm T, V^0}$) within charged jets for different cone radii compared to the inclusive ratio (black filled circles). Data is shown as points while lines show results from MC generators.}
    \end{center}
\end{figure}

The measurements presented in the previous paragraphs have shown that there are no significant modifications of particle production at high $\pt$ as compared to pp collisions while at low $\pt$ spectra are modified and long-range correlation structures appear in high-multiplicity events. Two-particle correlations can assess if jet yields in this low-$\pt$ region are modified~\cite{milano,1406.5463}. Figure~\ref{minijets_lk0ratio} (left panel) shows the near-side jet-like associated yield as a function of forward multiplicity after subtraction of the long-range contribution estimated at large pseudorapidity differences. The yield is basically independent of multiplicity except at low multiplicity where a reduction is observed.
This observation is similar on the away side. This finding is consistent with a picture in which the minijet yields in p--Pb collisions stem from a superposition of nucleon--nucleon collisions with incoherent fragmentation, while the long-range double-ridge correlation appears unrelated to minijet production. The reduction at low multiplicities may be related to event selection biases, see Section~\ref{sec_centrality}.

\subsection{Particle Content of Jets}

In Pb--Pb collisions, the inclusive $\Lambda/$K$^{\rm 0}_{\rm S}$ ratio is significantly larger in central collisions than in peripheral collisions which is interpreted as a sign of the partonic degrees of freedom and the collective expansion in a quark--gluon plasma. In p--Pb collisions, the $\Lambda/$K$^{\rm 0}_{\rm S}$ ratio is also larger in high-multiplicity than in low-multiplicity collisions~\cite{Abelev:2013haa}.
The track-based reconstruction of jets allows us to study if this enhancement is also present within jets~\cite{zhang}. Figure~\ref{minijets_lk0ratio} (right panel) presents the $\Lambda/$K$^{\rm 0}_{\rm S}$ ratio within charged jets for jet $\pt$ above \unit[10]{\gevc} for high-multiplicity events. This ratio does not exceed about 0.3 within the jet compared to about 0.7 for the inclusive ratio at \unit[3--4]{\gevc}. These low values are consistent with results from PYTHIA pp collisions and low-multiplicity p--Pb collisions showing that the fragmentation seems to occur in a similar way in p--Pb and pp collisions.

\subsection{Centrality Determination in p--Pb Collisions}
\label{sec_centrality}

The division of events into centrality classes with the aim of measuring observables as a function of impact parameter is not straightforward in p--Pb collisions as several biases occur~\cite{toia}:
the number of nucleons participating in the collision is rather modest, leading to a large influence of fluctuations;
due to the lower multiplicity, event selection biases play a larger role in p--Pb than in Pb--Pb collisions (such as the suppression of events with jets when low-multiplicity events are selected);
furthermore, in peripheral collisions geometric biases are present which reduce the number of parton--parton interactions per nucleon--nucleon collision.

\begin{figure}[t!]
    \begin{center}
        \includegraphics*[width=0.49\textwidth]{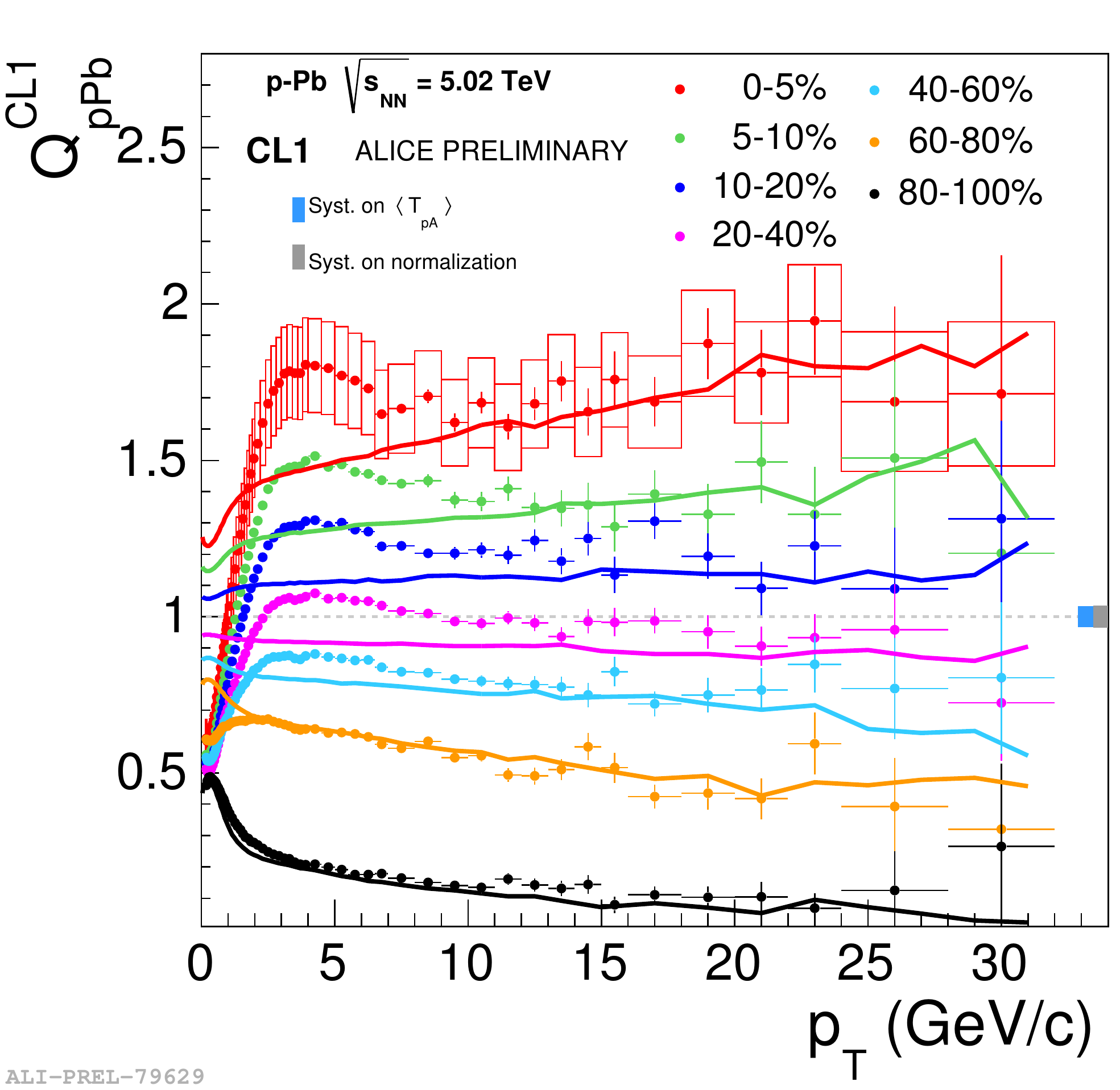}
        \hfill
        \includegraphics*[width=0.49\textwidth]{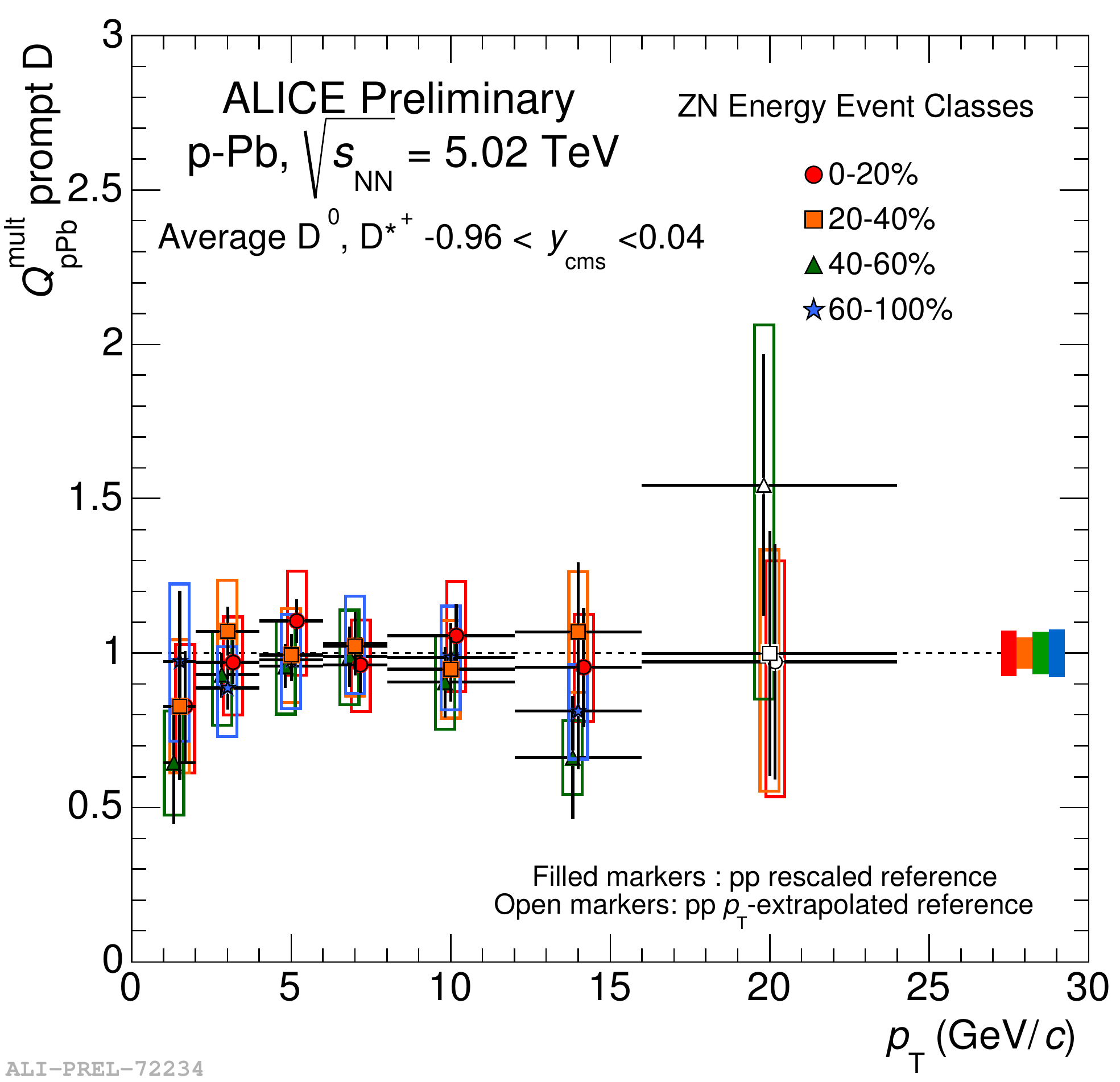}
        \caption{\label{centrality} Left panel: $Q_{\rm pPb}$ as a function of $\pt$ for charged particles using a mid-rapidity detector (denoted with CL1) to slice the events in classes. The data is shown as points while the G-PYTHIA model (see text) is shown as lines. The systematic error on the spectra is only shown for
        the 0-5\% bin and is the same for all others.
        \newline Right panel: $Q_{\rm pPb}$ as a function of $\pt$ for D mesons using the hybrid centrality method, for details see text.}
    \end{center}
\end{figure}

\begin{figure}[b!]
    \begin{center}
        \includegraphics*[width=0.49\textwidth]{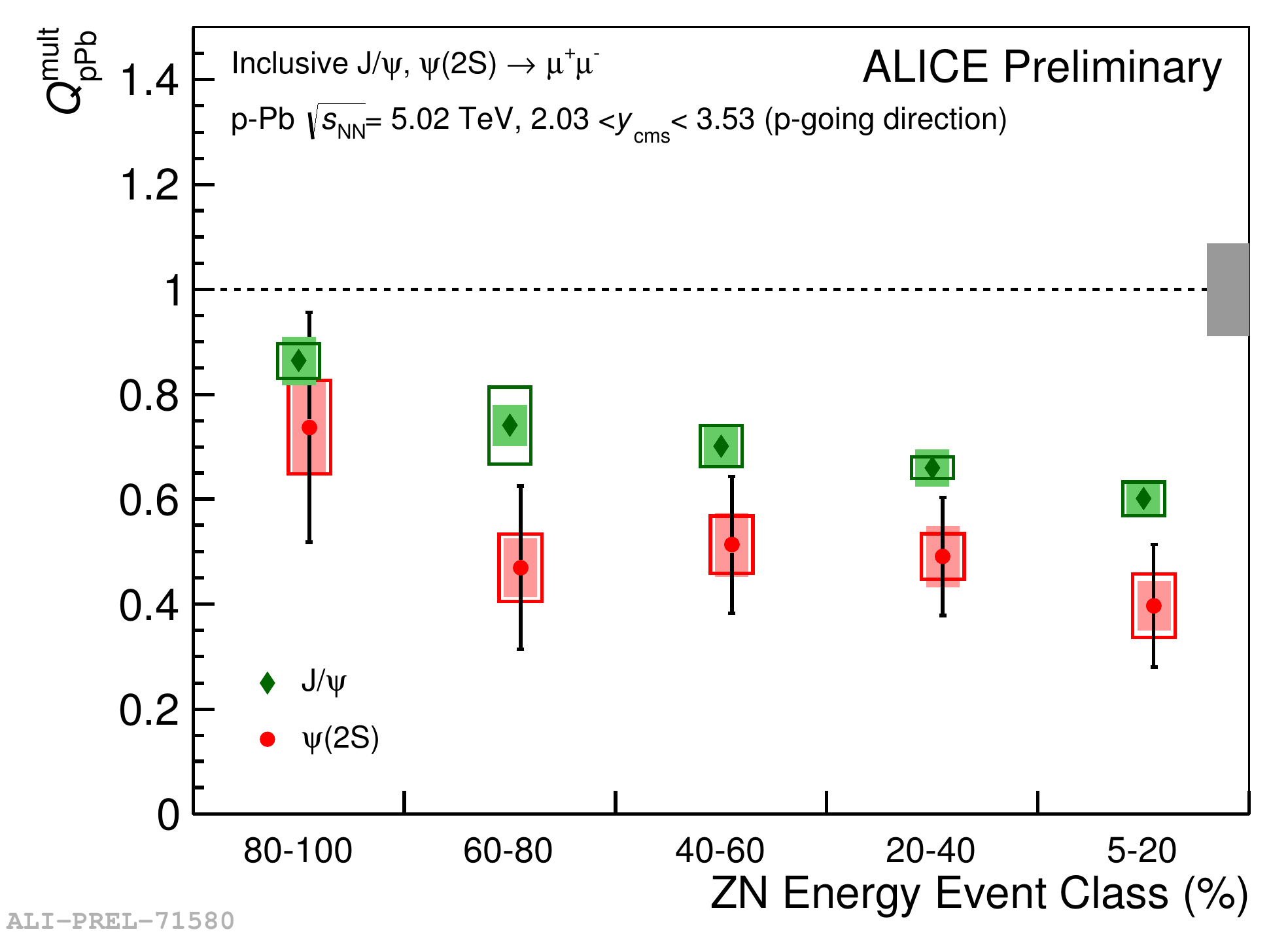}
        \hfill
        \includegraphics*[width=0.49\textwidth]{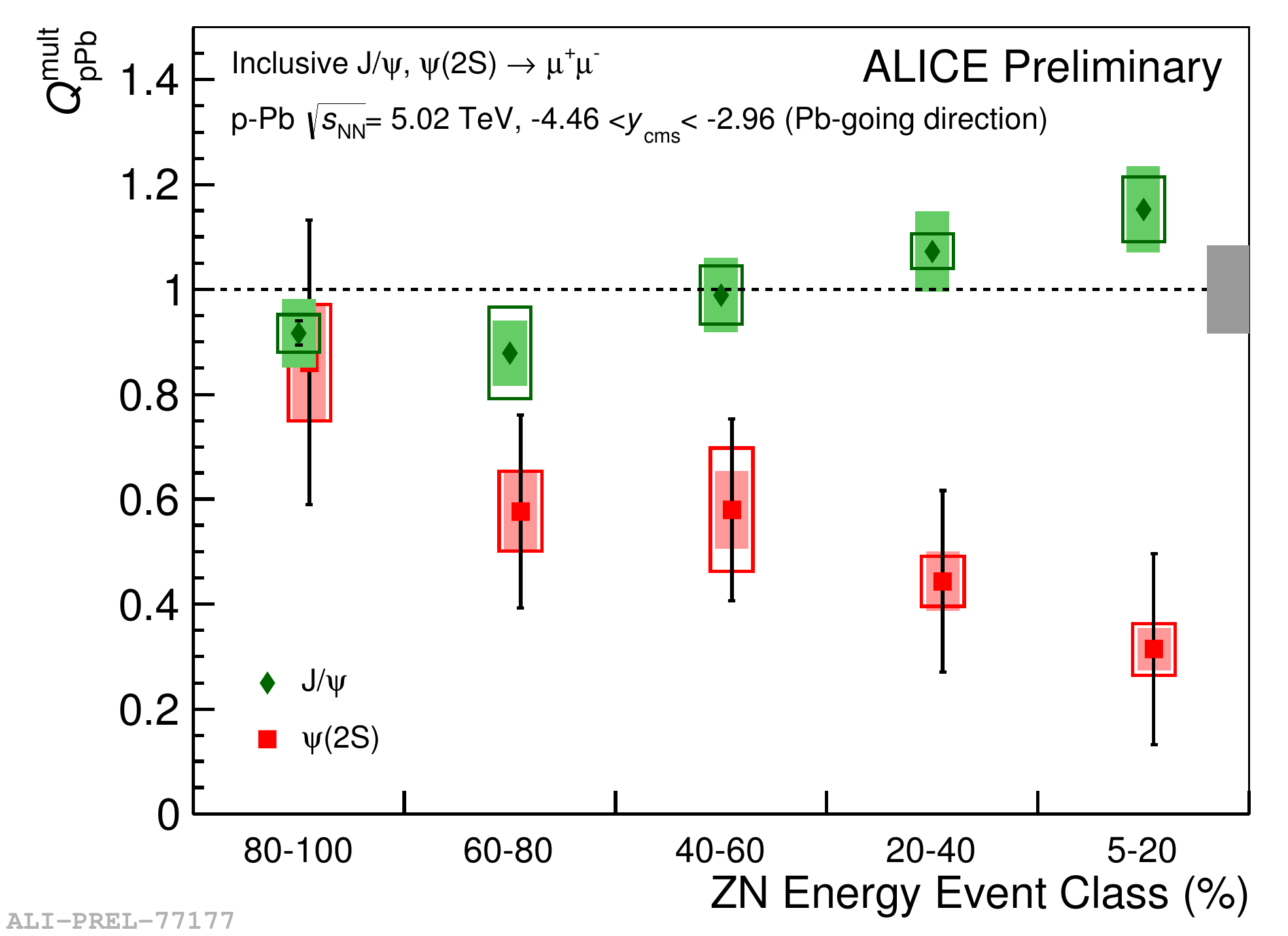}
        \caption{\label{charmonia} Inclusive $Q_{\rm pPb}$ for J/$\psi$ (green diamonds) and $\psi$(2S) (red circles) as a function of event activity class in p-going direction (left panel) and Pb-going direction (right panel). }
    \end{center}
\end{figure}

These biases can be illustrated with a simple model where a Glauber calculation of the number of nucleon--nucleon collisions $N_{\rm coll}$ is coupled with PYTHIA (called G-PYTHIA); for each nucleon--nucleon collision a PYTHIA pp minimum-bias collision is generated. Subsequently, the events are divided into classes based on the multiplicity at mid-rapidity and the ``standard'' nuclear-modification factor is calculated with respect to minimum-bias PYTHIA events. $N_{\rm coll}$ is determined based on the distribution of the event multiplicity. To underline that this observable is biased we call it $Q_{\rm pPb}$ instead of the ``standard'' $R_{\rm pPb}$. A value close to unity is expected for such an incoherent superposition.
Figure~\ref{centrality} (left panel) shows the so-obtained $Q_{\rm pPb}$ compared to the measured quantity.
Surprisingly, the $Q_{\rm pPb}$ of G-PYTHIA is between 0.1 and 1.8 at high $\pt$ depending on the event class and shows a $\pt$ dependence contrary to the expectation of unity. The data agrees overall well with this simple model, in particular the large difference between the different event classes at high $\pt$ as well as the slopes are reproduced showing the significant influence of the event selection.

A hybrid method is proposed with the aim of reducing the discussed biases. In this method events are sliced into classes based on the signal in the zero-degree neutron detectors (ZN), introducing a maximal pseudorapidity gap between the detector used for the event selection and the detectors used to measure the event activity under study. Then, the average number of nucleon--nucleon collisions in a class is obtained by assuming a scaling of the multiplicity with $N_{\rm part}$ (the number of nucleons participating in the collision). The resulting $Q_{\rm pPb}$ are consistent with unity at high-$\pt$ for all event classes. The hybrid method allows us to compute the $Q_{\rm pPb}$ for other particles as a function of event activity. As an example, Figure~\ref{centrality} (right panel) presents the $Q_{\rm pPb}$ of D mesons which shows no significant dependence on event class: the production of D mesons scales with $N_{\rm coll}$.

\subsection{Surprising Nuclear Modification Factor of J/$\psi$ and $\psi$(2S)}

The nuclear modification factor of inclusive J/$\psi$ and $\psi$(2S) production is found to show an interesting pattern. Figure~\ref{charmonia} presents $Q_{\rm pPb}$ as a function of event class for the p-going and Pb-going directions \cite{blanco,lakomov,arnaldi}. Both have been measured in the dimuon channel at forward rapidities. The J/$\psi$ shows a suppression in the p-going direction, which increases with event activity, while no significant modification is observed in the Pb-going direction. These findings are consistent with the expectation from the nuclear modification of the parton distribution functions (shadowing). The $\psi$(2S) shows a similar suppression pattern as the J/$\psi$ in the p-going direction. In the Pb-going direction, however, contrary to the J/$\psi$, the $\psi$(2S) is also suppressed. The expectation from shadowing though is that both should show a similar suppression, suggesting that additional, e.g. final state, effects are affecting the observed yields.

\section{Results from Pb--Pb Collisions}

As discussed in previous Quark Matter conferences, the hot and dense medium produced at LHC is larger and hotter than observed previously at lower-energy collisions, which results in stronger jet quenching and medium effects. Newly presented results quantify these effects with high precision and for numerous particle species.

\subsection{Jet Quenching in the Heavy-Flavour Sector}

The study of heavy-flavour decays allows the measurement of the $R_{\rm AA}$ of combined D$^{\rm 0}$, D$^+$ and D$^{*+}$~\cite{festanti} and their elliptic-flow coefficient $v_2$ \cite{bailhache,dv2}. The fact that the $R_{\rm AA}$ is significantly below 1 at high $\pt$ and that a non-zero $v_2$ is measured shows the sizable interactions of $c$ quarks and D mesons with the hot and dense medium. These high-precision $R_{\rm AA}$ and $v_2$ measurements provide significant constraints to models since both have to be reproduced simultaneously.

\subsection{Elliptic Flow}

\begin{figure}
    \begin{center}
        \includegraphics*[width=0.49\textwidth,clip=true,trim=0 0 45 0]{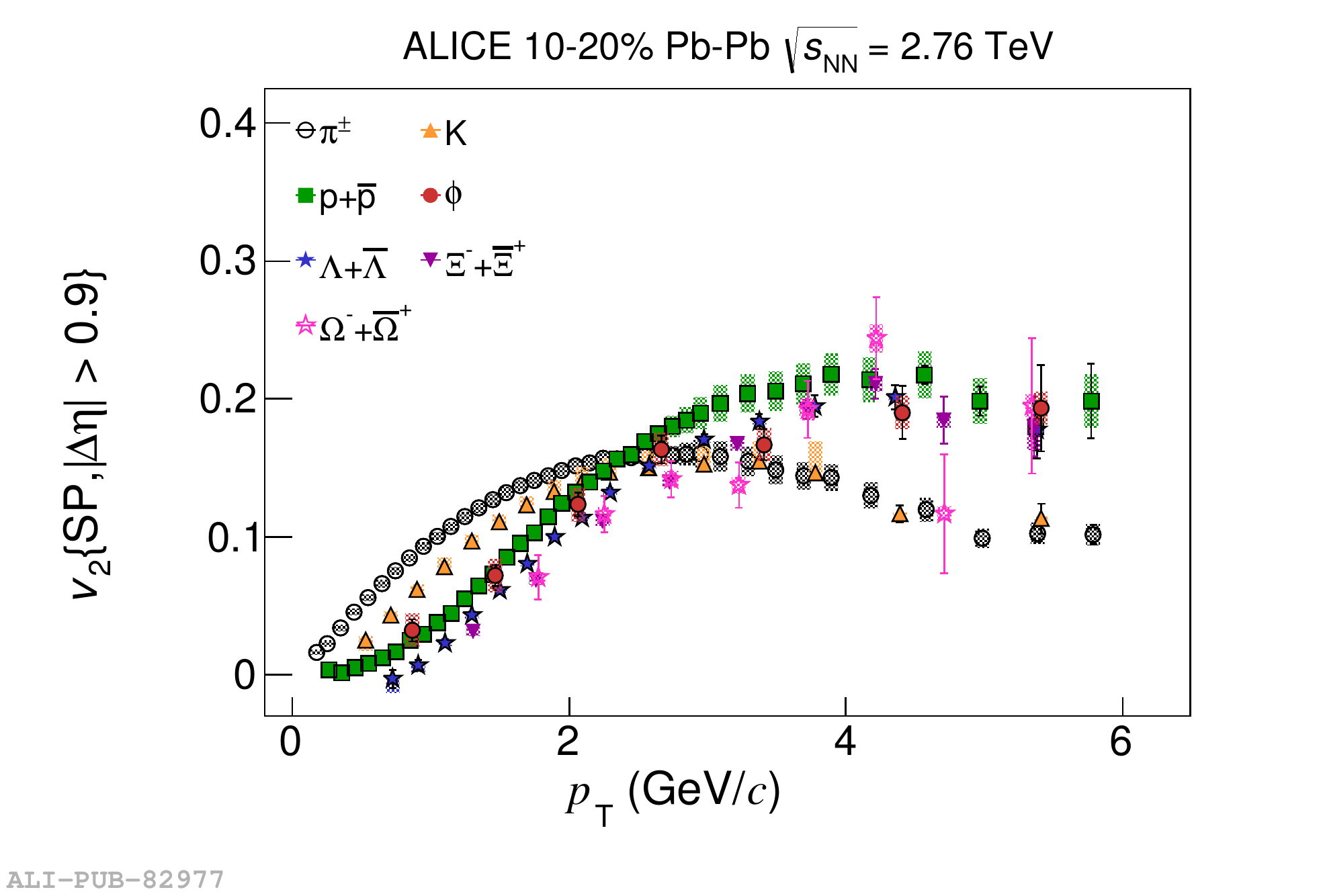}
        \hfill
        \includegraphics*[width=0.49\textwidth,clip=true,trim=0 0 45 0]{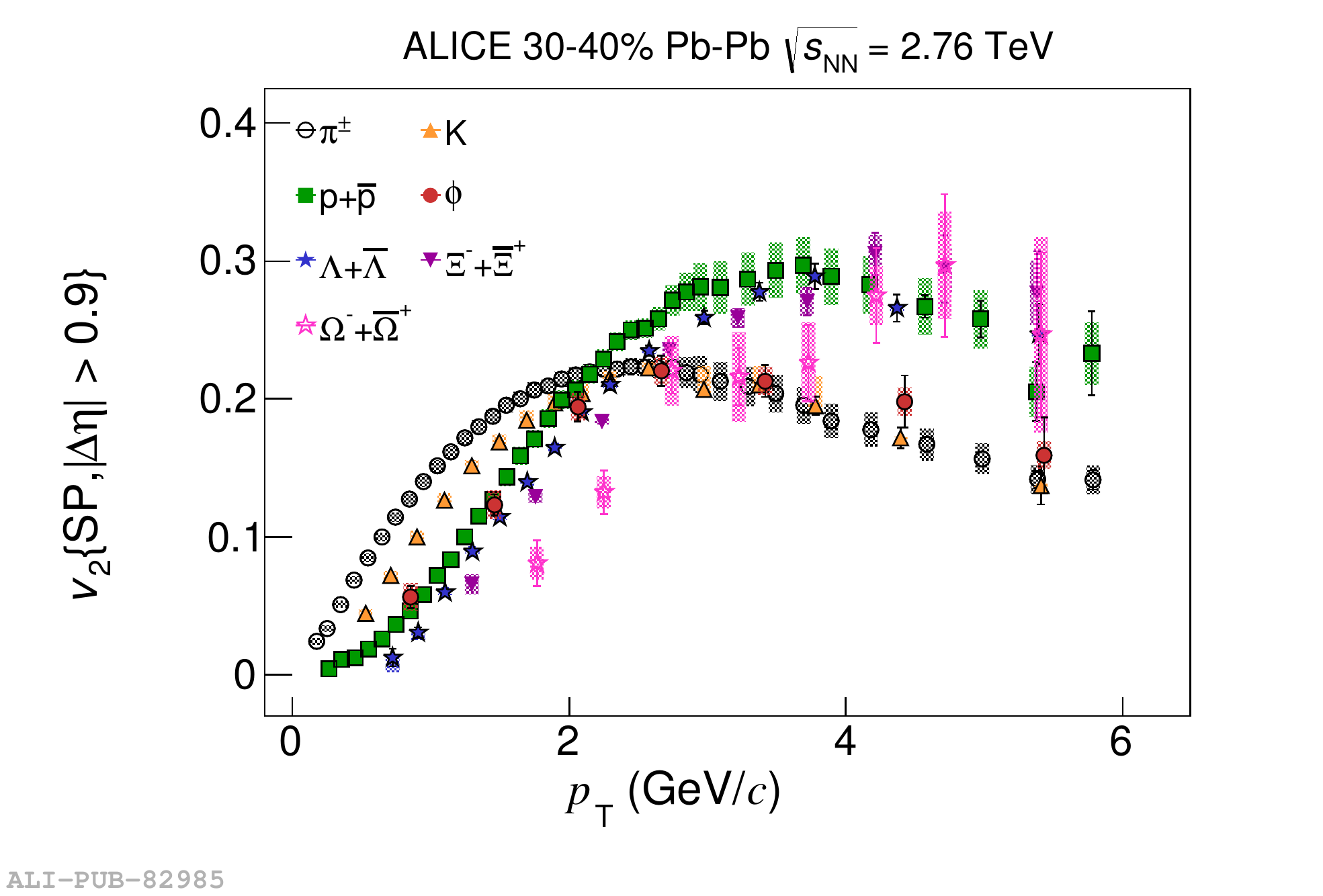}
        \caption{\label{v2} Elliptic-flow coefficient $v_2\{{\rm SP}\}$ as a function of $\pt$ for $\pi$ (black open circles), K (orange upward-pointing triangles), p+$\bar{\rm p}$ (green squares), $\phi$ (red filled circles), $\Lambda$ (blue filled stars), $\Xi$ (violet downward-pointing triangles) and $\Omega$ (pink open stars) measured in central 10--20\% (left panel) and 30--40\% (right panel) Pb--Pb collisions \cite{1405.4632}.}
    \end{center}
\end{figure}

Elliptic flow has been measured as a function of $\pt$ and centrality for eight different particle species: $\pi$, K$^\pm$, K$^{\rm 0}_{\rm S}$, p+$\bar{\rm p}$, $\phi$, $\Lambda$, $\Xi$ and $\Omega$ \cite{dobrin,1405.4632}. Figure~\ref{v2} presents the $v_2$ extracted with the scalar-product method for two centrality classes. The measured $v_2$ values of the different particles are found to be ordered according to their particle masses for $\pt < \unit[2.5]{\gevc}$. Comparisons to hydrodynamic calculations with VISHNU~\cite{vishnu} show a good agreement for $\pi$ and K while the p $v_2$ is underestimated and that of $\Lambda$ and $\Xi$ are overestimated. Dividing the obtained $v_2$ as well as the $\pt$ by the number of constituent quarks ($n_{\rm q}$) allows us to assess the so-called NCQ scaling~\cite{ncq}. This scaling, where $v_2/n_{\rm q}$ as a function of $\pt/n_{\rm q}$ is expected to be universal for all particle species, is violated by about 20\% in central collisions.

\subsection{The $\phi$ Meson}

\begin{figure}[t!]
    \begin{center}
        \hspace{0.2cm}
        \includegraphics*[width=0.45\textwidth]{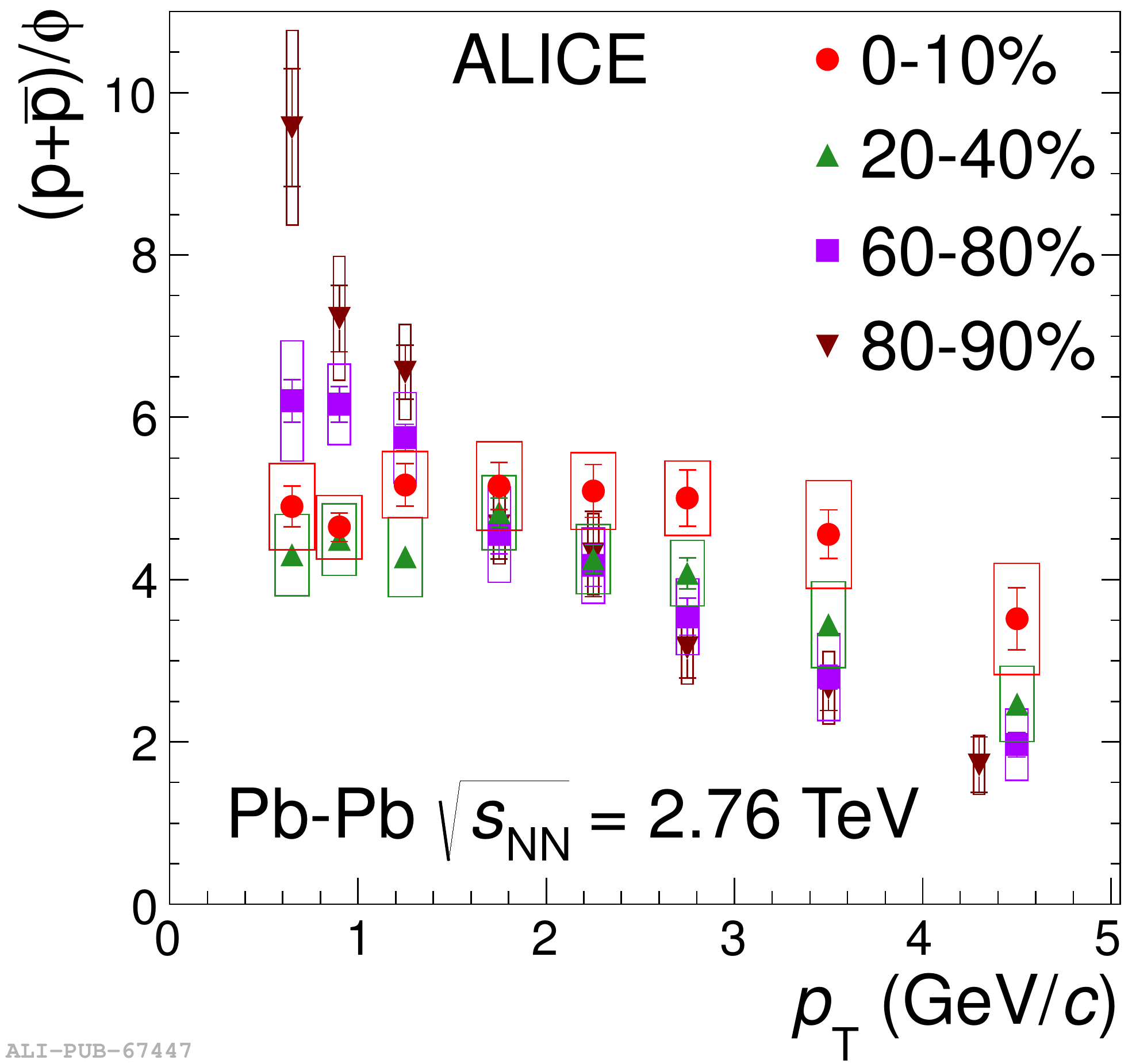}
        \hfill
        \includegraphics*[width=0.45\textwidth]{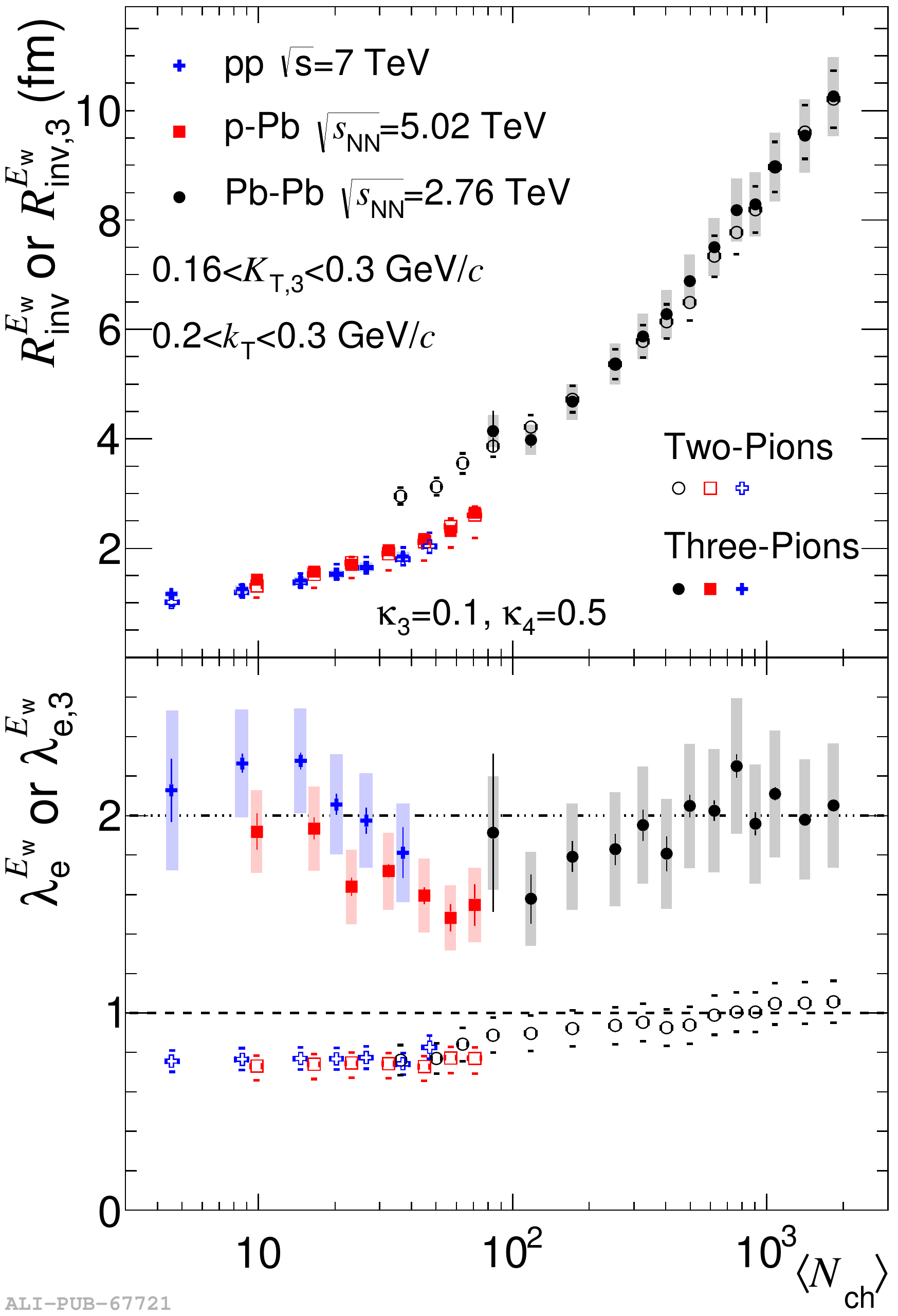}
        \hspace{0.2cm}
        \caption{\label{pphi_femto} Left panel: Proton to $\phi$ ratio as a function of $\pt$ for different Pb--Pb centrality classes~\cite{1404.0495}. \newline
        Right panel: Femtoscopic radii extracted from two- and three-pion cumulants together with the associated $\lambda$ parameters~\cite{1404.1194}.}
    \end{center}
\end{figure}

The $v_2$ of the $\phi$ shows an interesting pattern: below a $\pt$ of \unit[2.5]{\gevc} it follows mass ordering while at higher $\pt$ the behavior depends on centrality: in central collisions, its $v_2$ is close to that of the p, while in mid-central collisions its $v_2$ is closer to the one of the $\pi$ (see Fig.~\ref{v2}). Comparing the $\pt$ spectra of $\phi$ and p reveals that they have a similar shape in central collisions up to about \unit[4]{\gevc} (left panel of Fig.~\ref{pphi_femto}) \cite{1404.0495,bellini}.
This is expected in a picture where the spectral shape is driven by radial flow.
Combining this finding with that for the $v_2$ suggests that the mass (and not the number of constituent quarks) drives $v_2$ and spectra in central Pb--Pb collisions for $\pt < \unit[4]{\gevc}$. It is interesting to note that also in p--Pb collisions the shape of the $\pt$ spectra of $\phi$ and p become more similar for high-multiplicity events~\cite{andrei}.

\subsection{Identified-Particle Spectra}

The ALICE collaboration has presented yields and spectra for 12 particle species ($\pi$, K$^\pm$, K$^*$, K$^0$, p, $\phi$, $\Lambda$, $\Xi$, $\Omega$, d, $^3$He, $^3_\Lambda$H) in up to 3 collision systems (and, for pp collisions, 3 different center of mass energies). In particular the measurement of the $\pt$ and centrality dependence of the d and the nuclei ($^3$He, $^3_\Lambda$H) spectra should be pointed out~\cite{martin}. It is interesting to note that the yields of d, $^3$He and $^3_\Lambda$H are correctly calculated in equilibrium thermal models.
Furthermore, the yields of multi-strange baryons have been measured as a function of event multiplicity showing a smooth evolution from pp over p--Pb to Pb--Pb collisions for the yield ratios to $\pi$ or p~\cite{alexandre}.
The large amount of data allows a stringent comparison to thermal models which describe particle production on a statistical basis~\cite{floris}.

\subsection{Source Sizes}

For the first time, femtoscopic radii were extracted with three-pion cumulants~\cite{1404.1194,gangadharan}. This approach reduces non-femtoscopic effects contributing to the extracted radii significantly. Figure~\ref{pphi_femto} (right panel) presents the extracted radii using an Edgeworth expansion~\cite{edgeworth} compared to results using two-pion correlations for pp, p--Pb and Pb--Pb collisions. The size of the system extracted in this way is 35--45\% larger in Pb--Pb collisions than in p--Pb collisions at the same multiplicity showing that there are differences between high-multiplicity p--Pb collisions and peripheral Pb--Pb collisions. On the contrary, the radii in p--Pb collisions are only 5--15\% larger than in pp collisions at the same multiplicity. The radii in pp and p--Pb collisions can be reproduced in a CGC initial-state model (IP-GLASMA) without a hydrodynamic phase~\cite{glasma} while such calculations underestimate the radii measured in Pb--Pb collisions.

\begin{figure}
    \begin{center}
        \includegraphics*[width=0.7\textwidth]{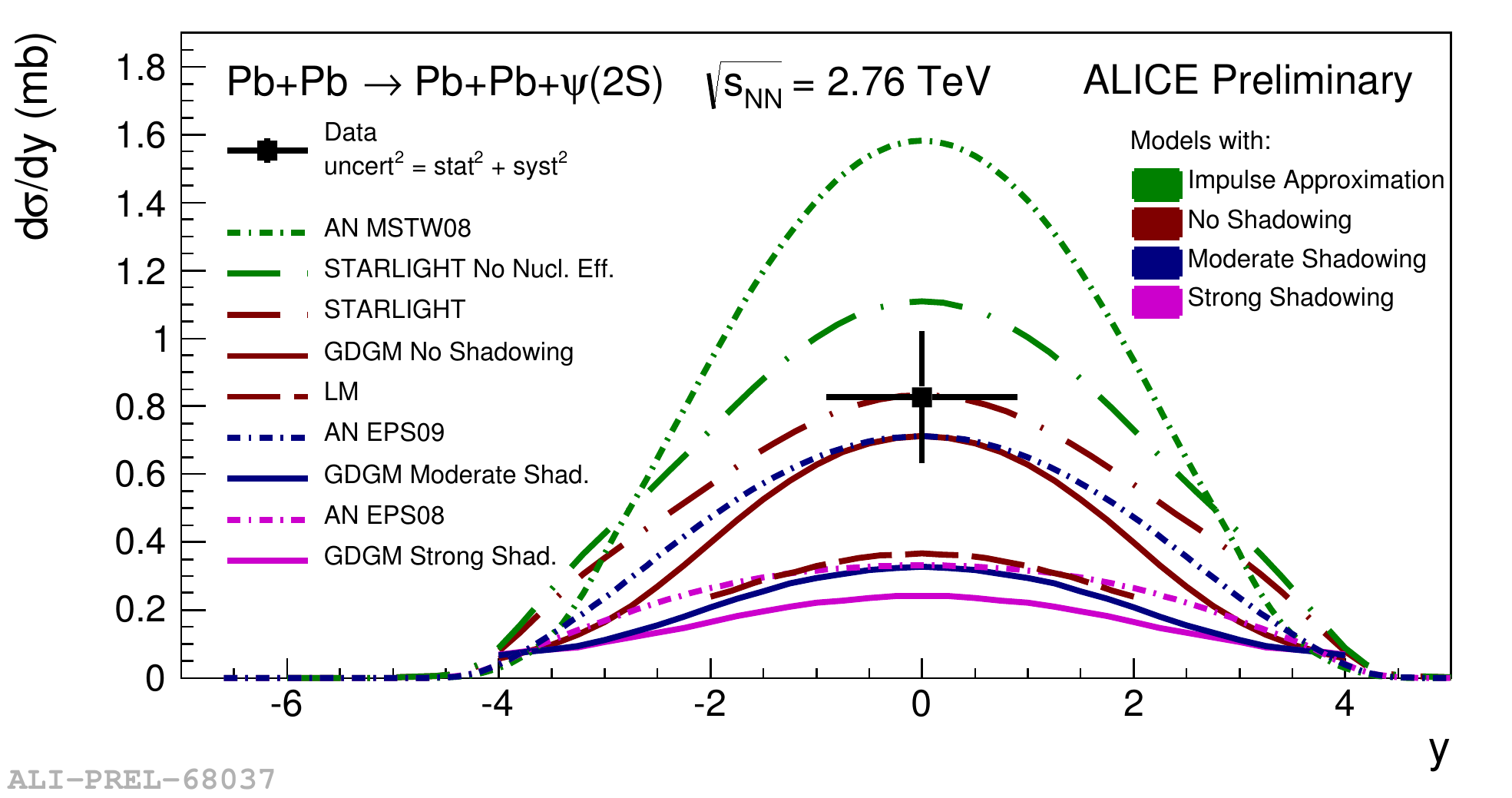}
        \caption{\label{upc} $d\sigma/dy$ for coherent $\psi$(2S) photoproduction measured in ultra-peripheral Pb--Pb collisions compared to model predictions.}
    \end{center}
\end{figure}

\subsection{Exclusive Production of the $\psi$(2S) State}

In ultra-peripheral Pb--Pb collisions particles can be produced exclusively. The first measurements of coherent $\rho$ and $\psi$(2S) photoproduction in a nuclear target have been performed~\cite{nystrand}. This required a special trigger and event selection where the decay products of the produced state are the only objects observed in the detector. For the $\psi$(2S), two different channels have been studied: the decay to two leptons as well as the decay to two leptons and two $\pi$. For each channel about 20 candidates were found. Figure~\ref{upc} shows the extracted cross section compared to model predictions. Despite the present uncertainties, the measurement allows us to conclude that models including strong shadowing as well as those without any nuclear effects are disfavored.

\section{Results from pp Collisions}

The ALICE collaboration has also presented several measurements in pp collisions.
In particular, there is an interesting measurement of the particle-type dependence of jet fragmentation functions for $\pi$, K and p in jets of \unit[5--20]{\gevc} and constituent $\pt$ between 0.2 and \unit[20]{\gevc}~\cite{lu}. Such measurements provide important constraints on fragmentation functions at low $\pt$.
Furthermore, a direct photon measurement has been presented for $1 < \pt <$~\unit[4]{\gevc} consistent with next-to-leading order pQCD~\cite{kohler}.

\section{Conclusions}

In Pb--Pb collisions, the presented results significantly improve the precision of previous measurements in various areas.
In particular, a measurement of elliptic flow with identified particles shows a clear mass ordering for light and strange hadrons for $\pt <$~\unit[2.5]{\gevc}. Spectra and $v_2$ measurements of the $\phi$ meson suggests that the mass (and not the number of constituent quarks) drives the spectral shape and the size of the elliptic flow in central collisions for $\pt <$~\unit[4]{\gevc}.
While there are several observables which are approximately consistent with a description of p--Pb collisions as incoherent superposition of nucleon--nucleon collisions at high $\pt$, some measurements hint to novel effects at low $\pt$ which are potentially of collective origin. Furthermore, the suppression pattern of $\psi$(2S) compared to J/$\psi$ indicates that significant final-state effects are at play. These findings still need to be reconciled theoretically and promise that p--Pb collisions will continue to be a very exciting field in the future.








\end{document}